# An Empirical Investigation of Cash Conversion Cycle of Manufacturing Firms and its Association with Firm Size and Profitability

- Nusrat Jahan[*]

**Abstract**
The purpose of this empirical study is to investigate Cash Conversion Cycle of thirty manufacturing firms listed in Dhaka Stock Exchanges under six different categories, which are, Food and allied, Pharmaceuticals and chemical, Cement, Textile, Engineering and Miscellaneous. This paper sets industry average Cash Conversion Cycle for these six industries and examines the relationship of Cash Conversion Cycle with firm size and profitability. This study did not find statistically significant differences among the Cash Conversion Cycle of varying manufacturing industries. The result of this study indicates a statistically significant negative relationship between the Cash Conversion Cycle and profitability, especially in terms of Return on Equity. The result also shows that the Cash Conversion Cycle of manufacturing firm also has significant negative relationship with firm size, when measured in terms of net sales. The present study contributes to the literature on working capital management written in the context of Bangladesh.
**Keywords**: Cash Conversion Cycle, size, profitability, manufacturing industry.

## 1. Introduction

Working capital management refers to the management of current assets and current liabilities of a firm in order to achieve a balance between profitability and risk that contributes positively to the value of a firm (Gitman, 2000, p.616). One of the popular and powerful measures of working capital management is Cash Conversion Cycle. Firm's typically follow a cycle, in which they purchase inventory, sell goods on credit and then collect accounts receivables. This cycle is referred to as Cash Conversion Cycle. Therefore, the Cash Conversion Cycle can be defined as the length of time between the firm's actual cash expenditures to pay for raw materials and its own cash receipts from the sale of finished goods.

[*]The author is Lecturer, School of Business, Independent University, Bangladesh. The views expressed in this article are the authors' own.

Thus the Cash Conversion Cycle equals the average length of time a dollar is tied up in current assets (Ehrhardt & Brigham, 2003, p.581). Therefore, it is evident that Cash Conversion Cycle focuses only on the time period for which cash flow is engaged in the cycle and does not consider the amount of fund committed to a product as it moves through the Cash Conversion Cycle (Nobanee, 2009).The Cash Conversion Cycle is a more powerful measure of working capital management and firm's liquidity, compared to the static traditional measures which are found to be inadequate and misleading in the evaluation of a firm's liquidity. According to Moss and Stine (1993) a useful way of assessing the liquidity of a firm is through Cash Conversion Cycle. The traditional measures of liquidity such as the current ratio and quick ratio are useful liquidity indicators but they focus on static balance sheet values (as cited in Uyar, 2009). On the contrary, Jose et al (1996) advocates Cash Conversion Cycle as a dynamic measure of ongoing liquidity management since it combines both balance sheet and income statement data to create a measure associated with time dimension (as cited in Uyar, 2009). The length of Cash Conversion Cycle is expected to vary across industries because there is likely to be an industry effect on an individual firm's Cash Conversion Cycle. Therefore, determination of industry average is significant since it allows the individual firms within an industry to evaluate its own performance relative to the industry and prevent itself from probable liquidity crisis. The primary focus of this empirical study is determining and investigating the differences in Cash Conversion Cycle of six different manufacturing industries of Bangladesh.

Furthermore, this study also investigates the relationship between Cash Conversion Cycle and size of the firm. Small firm may not have much investment in fixed assets. But it generally has a higher level of investment in current assets since these firms faces severe problem in collecting their debts, especially in Bangladesh. Besides, the role of current liabilities in financing current assets is far more significant in case of small firms. Because, unlike large firms, small firms face difficulty in raising long-term finances (Pandey, 2005, p. 820). Thus, Cash Conversion Cycle is likely to have a negative relationship with firm size. Earlier studies also have proved that size of firm is one of the variables that affect Cash Conversion Cycle. Therefore, this study aims to establish association between Cash Conversion Cycle and firm size in the context of manufacturing sector of Bangladesh. A long Cash Conversion Cycle might increase profitability because it leads to higher sales, greater investment in inventories and trade credit granted by the firm. On the contrary, shorter Cash Conversion Cycle harms firms' profitability. However, corporate profitability might decrease with the Cash Conversion Cycle, if the cost of investment in working capital rises faster than the benefits of holding more inventories and granting more trade credit to customers (Gill, Bigger & Mathur, 2010). Therefore, Cash Conversion Cycle may have both positive and negative impact on the firm's profitability. Earlier studies have also established both positive and negative relationship between Cash Conversion Cycle and profitability. Thus, the present study also attempts to determine the association between Cash Conversion Cycle and profitability of manufacturing firms of Bangladesh.

## 2. Literature Review

The corporate finance literature written in the context of Bangladesh has traditionally focused on capital structure, investments, dividends and firm's valuations. The study of working capital management in the context of Bangladesh is very little explored by the researchers. Therefore, considering this research gap, the present empirical study seeks to shed light onto this area by evaluating one of the powerful measures of working capital management, which is Cash Conversion Cycle.

Working capital management is concerned with the management of current assets and current liabilities and the interrelationship between them. Its operational goal is to manage the current assets and current liabilities in such as way that a satisfactory level of working capital is maintained (Khan and Jain, 2007, p. 13). One of the powerful measures of working capital management is Cash Conversion Cycle. Schilling (1996) and Gallinger (1997) advocate Cash Conversion Cycle as a powerful working capital evaluation technique. The advantage of this technique is that it can be used to evaluate changes in circulating capital and thereby facilitate the monitoring and controlling of its components (as cited in Lyroudi and Lazaridis, 2000). According to Richards and Laughlin (1980) Cash Conversion Cycle provides dynamic insights compared to traditional static liquidity ratios. They suggested a positive relationship between the current and quick ratio and the Cash Conversion Cycle (as cited in Lyroudi and Lazaridis, 2000). Mose and Stine (1993) and Jose, Lancaster and Stevens (1996) suggest that traditional static measures focus on a single point in time whereas Cash Conversion Cycle, being a dynamic measures of time, considers the time it takes for a firm to go from cash outflow to cash inflow (cited in Uyar, 2009).

The Cash Conversion Cycle components are inventory conversion period, receivables collection period and payable deferral period. The inventory conversion period is the number of days it takes to convert raw materials into finished goods and sell those goods to the firm's customers. The receivables collection period is the time taken to collect receivables from the customers. The payable deferral period refers to the time taken by the firm to pay its own obligations. Cash Conversion Cycle reports the result in length of time, which is important because the amount of capital needed to finance the company is related to the speed with which 'input' is converted into 'output' and payment is received for the sales of this 'output'. The following formula is used to calculate the Cash Conversion Cycle of a firm (Kaen, 1995, P. 731):

Cash Conversion Cycle = (Inventory conversion Period + Receivables collection period – Payable deferral period)

$$\text{Inventory Conversion Period} = \frac{\text{Inventory}}{\text{Cost of Goods Sold} / 365}$$

$$\text{Receivables Collection Period} = \frac{\text{Account Receivables}}{\text{Sales} / 365}$$

$$\text{Payable Deferral Period} = \frac{\text{Accounts Payable}}{\text{Cost of Goods Sold} / 365}$$

Cash Conversion Cycle can be positive as well as negative. A positive result indicates the number of days a company must borrow or tie up capital while awaiting payment from a customer. According to Hutchison, Farris and Andres (2007) a negative Cash Conversion Cycle indicates the number of days a company has received cash from sales before it must pay its suppliers (cited in Uyar, 2009). The main goal of any firm would be to shorten its Cash Conversion Cycle. However, the changes in the length of the Cash Conversion Cycle are likely to have costs as well as benefits for the firm. Level of working capital and Cash Conversion Cycle of a firm is affected by a variety of factors and one of the important factors is nature of the business. According to Belt (1985) retailing and wholesaling firms both have Cash Conversion Cycle shorter than those of manufacturing firms (as cited in Lyroudi and Lazaridis, 2000). Studies by Hawawini et al (1986), Weinraub and Visscler (1998), Wu (2001) and Filbeck (2005) have showed an industry effect on firm's working capital policies which was explained by differences in trade credit and investment in inventories across industries (as cited in Pedro et al.,2009). Besides Smith (1987) and Ng, Smith and Smith (1999) suggested a wide variation in credit terms across industries but little variation within industries Nisken and Nisaken (2006) also found differences in levels of accounts receivables and accounts payables between industries (as cited in Pedro et al.,2009). Furthermore, Hutchison et al (2007) advocates that analysis of an individual firm's Cash Conversion Cycle is helpful while industry benchmarks are crucial for a company to evaluate its Cash Conversion Cycle's performance and assess opportunities for improvement (as cited in Uyar, 2009). The present study also aims at investigating the differences in Cash Conversion Cycle of different industries of Bangladesh.

Empirical evidences suggests that size is another variable that affects working capital management, hence the Cash Conversion Cycle of a firm. Moss and Stine (1993) examined the relationship between the Cash Conversion Cycle and the size of US retail firms and found that larger firms have shorter Cash Conversion Cycle, which implies smaller companies should try to better manage their Cash Conversion Cycle (as cite in Lyroudi and Lazaridis, 2000). Kieschnick, LaPlante and Moussawi (2006) and Chiou, Cheng and Wu (2006) demonstrated a positive relationship between size and Cash Conversion Cycle (as cited in Pedro et al.,2009). It is expected in the present study, as in previous studies by Moss and Stine (1993), that size will also negatively influence the Cash Conversion Cycle of manufacturing firms of Bangladesh.

Level of profit of a firm is also affected by its Cash Conversion Cycle. Sanger (2001) advocates that working capital, though it represents a safety cushion for providers of short-term funds of the company, however, from operating point of view, excessive level of working capital is looked at as a restraint on financial performance, since these assets do not contribute to return on equity (as cited in Eljelly, 2004). Deloof (2003) found a significant negative correlation between gross operating income and the number of days in account receivables, inventories and accounts payables of Belgian firms. Raheman and Nasr (2007) established a strong negative relationship between variables of working capital management and profitability of the firm (as cite in Gill et al., 2010). The present study also aims at investigating the relationship between Cash Conversion Cycle and profitability of manufacturing firms of Bangladesh.

## 3. Research Objectives

The objectives of this study are stated below:
1) Determination and examination of difference in Cash Conversion Cycle of six manufacturing industries, which are categorized as Food and Allied, Pharmaceutical and Chemical, Textile, Cement, Engineering and Miscellaneous.
2) Determining firm size in each industry and examining its association with Cash Conversion Cycle.
3) Determining profitability of firm in each industry and examining its association with Cash Conversion Cycle.

To fulfill the aforesaid objectives the following null-hypotheses are formulated:
1) The Cash Conversion Cycle of all six manufacturing industries is equal.
2) There is no linear relationship between Cash Conversion Cycle and firm size.
3) There is no linear relationship between Cash Conversion Cycle and firm's profitability.

## 4. Research Methodology

The purpose of this empirical study is to determine whether the Cash Conversion Cycle varies from industry to industry. It also aims to establish the relationship of Cash Conversion Cycle with firm size and profitability. This research primarily focuses on secondary data, which were obtained from the annual reports of the companies listed on the Dhaka Stock Exchange. The sample consists of thirty manufacturing companies from six different manufacturing industries, which are categorized in the Dhaka Stock Exchange as Engineering, Textile, Food and Allied, Pharmaceuticals and Chemical, Cement and Miscellaneous. For selection of sample, stratified random sampling method is used and 5 sample companies are selected under each type of industry. Service rendering companies were not included since they do not fit with the scope of the study. The data have been obtained for the year 2008 and includes yearly sales, cost of goods sold, receivables, payables, inventory, total asset, total equity and net profit of thirty sample companies. These data were used to calculate the Cash Conversion Cycle, average firm size and average profitability of firms' in each of the six different industries.

When the means of more than two groups are to be compared, one-way ANOVA is the appropriate statistical tool (Zikmund, 2000). Therefore, in order to determine, whether statistically significant differences exist among the Cash Conversion Cycle of six manufacturing industries, one-way ANOVA analysis was conducted. Here, the independent variable is industry which has six different levels. The dependent variable is Cash Conversion Cycle. Since there are six groups or levels, a t-test cannot be used for the testing of statistical significance.

In this study, the firm size is measured by total assets and net sales and profitability is measured by ROA and ROE. The most popular technique that indicates the relationship of one variable to another is simple correlation analysis (Zikmund, 2000). Therefore, to determine the linear relationship among variables, which are Cash Conversion Cycle, total assets, net sales, ROA and ROE, Pearson's Correlation analysis is applied. Furthermore, to determine the significance of a correlation coefficient, a t-test is performed, which hypothesizes that linear relationship between two variables is zero. The result of correlation and t-test is reported through a correlation matrix table with a footnote indicating each statistically significant coefficient. Data are summarized by Microsoft Excel and different tests were conducted by using SPSS.

## 5. Research Findings
### 5.1. Descriptive Statistics

Table 1 reports the descriptive statistics of the main variables used in this study. This table shows the yearly average values of inventory, accounts payables, accounts receivables, total assets, net sales, ROA, ROE and Cash Conversion Cycle of six different manufacturing industries. The reported differences in Cash Conversion Cycle of industries support the argument that there is an industry affect on the firms' Cash Conversion Cycle. Among all six industries, shown in the Table 1, Engineering has the longest Cash Conversion Cycle, followed by Textile and Cement industry.

Table 1: Results of Descriptive Statistics

| Industry | Inventory Yearly Average Taka | Receivable Yearly Average Taka | Payables Yearly Average Taka | CCC Yearly Average Days | Total Assets Yearly Average Taka | Net Sales Yearly Average Taka | ROA Yearly Average | ROE Yearly Average |
|---|---|---|---|---|---|---|---|---|
| **Engineering** | 296958687 | 296384226 | 63374288.8 | 316.84 | 1269076636 | 812443043.4 | 0.0037 | (0.00106) |
| **Textile** | 528611683 | 465890743 | 499359336.2 | 150.07 | 2183356231 | 1358315306 | 0.02345 | 0.04803 |
| **Pharmaceut-icals & Chemical** | 890655492 | 176938412 | 67296669.8 | 139.55 | 6159757151 | 3304305714 | 0.08588 | 0.18467 |
| **Cement** | 267411400 | 240915372 | 209935314.2 | 149.66 | 4548627904 | 2298462381 | (0.04091) | (0.05567) |
| **Food & Allied** | 53747687 | 46768527 | 123328069 | 74.13 | 2080697617 | 2858563395 | 0.01201 | 0.06152 |
| **Miscellaneo-us** | 125069011 | 22478128.2 | 2643588 | 141.89 | 679457272 | 496384916.6 | 0.044289 | 0.093012 |

**Source**: Annual Reports, concerned six industries

The reason behind these industries having longer Cash Conversion Cycle is that these enterprises make fairly large amount of investment in inventories and receivables to support their production, purchase and sales activity. Besides, these sectors tend to store inventory for longer period of time and production process also lengthen days in inventory. Furthermore, they also take most time to collect payment from its customers. Hence, these industries face more need to finance their working capital requirement. The lowest average Cash Conversion Cycle is found in the Food and Allied industry, because this sector stores inventory for the shortest period of time.

Table 2: Results of One-way ANOVA Analysis

|  | Sum of Squares | Degree of Freedom | Mean Square | F (Calculated) | Level of Significance | Critical value of F |
|---|---|---|---|---|---|---|
| **Between Groups** | 164,499.63 | 5 | 32,899.926 | 1.348 | 0.05 | 2.62 |
| **Within Groups** | 585,791.034 | 24 | 24,407.96 |  |  |  |
| **Total** | 750,290.663 | 29 |  |  |  |  |

**Source**: Self Computed

A formal test known as one-way ANOVA is introduced to compare the average Cash Conversion Cycle of six different manufacturing industries. This test allows to determine whether the differences in Cash Conversion Cycle are statistically significant or not. The test result shown in Table 2 reports that the critical value of F at the 0.05 level of significance for five and twenty four degrees of freedom is 2.62. Table 2 also indicates that the calculated value of F is 1.348 which is below the critical value 2.62. Therefore, the null hypotheses must be accepted. The test suggests that all the six different industries have approximately the same average Cash Conversion Cycle. Therefore, the differences that exist among Cash Conversion Cycle of manufacturing industries of Bangladesh are not statistically significant.

**Table 3: Correlation Matrix**

|  | CCC | Net Sales | Total Assets | ROA | ROE |
|---|---|---|---|---|---|
| CCC | 1 |  |  |  |  |
| Sig. (two-tailed) |  |  |  |  |  |
| N | 30 |  |  |  |  |
| Net Sales | (0.551)* | 1 |  |  |  |
| Sig. (two-tailed) | (0.349) |  |  |  |  |
| N | 30 | 30 |  |  |  |
| Total Assets | (0.259) | 0.811 * | 1 |  |  |
| Sig. (two-tailed) | (1.42) | 7.34 |  |  |  |
| N | 30 | 30 | 30 |  |  |
| ROA | (0.295) | 0.177 | 0.206 | 1 |  |
| Sig. (two-tailed) | (1.63) | 0.95 | 1.11 |  |  |
| N | 30 | 30 | 30 | 30 |  |
| ROE | (0.364)* | 0.333** | 0.286 | 0.980* | 1 |
| Sig. (two-tailed) | (2.07) | 1.87 | 1.58 | 26.06 |  |
| N | 30 | 30 | 30 | 30 | 30 |

\* Correlation is significant at the 0.05 level of significance (Sig. two-tailed) and 28 degree of freedom.
\*\* Correlation is significant at the 0.10 level of significance (Sig. two-tailed) and 28 degrees of freedom.

### 5.2. Cash Conversion Cycle and Firm Size

Second objective of this study is to investigate the relationship between Cash Conversion Cycle and firm size, for which two measures are used, which are net sales and total assets. Pearson Correlation analysis and t-test was performed to determine statistically significant relationship. Result of correlation analysis in Table 3 indicates that the linear relationship of Cash Conversion Cycle with total asset and net sales is negative. The correlation matrix shown in table 3 also points out that the negative correlation between Cash Conversion Cycle and net sales is statistically significant. The critical value of t-test at 0.05 level of significance (two-tailed) is 2.048 and the calculated value of 't' for correlation of Cash Conversion Cycle with total assets and net sales are -1.42 and -3.49 respectively. Since the value of 't' for correlation between net sales and Cash Conversion Cycle, exceeds the critical value, the null hypothesis should be rejected. Therefore, it can be concluded that there exist significant negative linear relationship between Cash Conversion Cycle and firm size, in the context of manufacturing firms of Bangladesh. This means, larger firms in Bangladesh have shorter Cash Conversion Cycle and smaller firms have longer Cash Conversion Cycle. The probable reason could be that, in Bangladesh smaller firms have less bargaining power, grants more trade credit to generate sales. Hence, it makes higher investment in inventories and also faces severe problem in collecting from their debtors. Therefore, the ratio of investment made by smaller firms in current assets is high compared to fixed assets, thus resulting in longer Cash Conversion Cycle. On the contrary, larger firms having more bargaining power tend to have less investment in current assets compared to smaller firms, thus resulting in shorter Cash Conversion Cycle. Therefore, small-scale firms should focus on shortening their Cash Conversion Cycle by reducing their inventory conversion period, receivables collection period and by increasing accounts payable period.

**5.3. Cash Conversion Cycle and Profitability of the Firm**

To determine the relationship of Cash Conversion Cycle with profitability two criteria has been used, which are Return on Equity (ROE) and Return on Assets (ROA). The correlation matrix (Table 3) shows that the linear relationship of Cash Conversion Cycle with ROA and ROE is negative indicating that longer Cash Conversion Cycle reduces profitability of the firm and vice-versa. Results of correlation analysis in Table 3 also indicate that the linear relationship of Cash

Conversion Cycle with ROE is statistically significant. The critical value of t-test at 0.05 level of significance (two-tailed) is 2.048 and the calculated value of 't' for correlation of Cash Conversion Cycle with ROE and ROA are -2.07 and − 1.63 respectively. Since the value of 't' for correlation between ROE and Cash Conversion Cycle, exceeds the critical value, the null hypothesis should be rejected. Therefore, the result suggests, there exist significant negative linear relationship between Cash Conversion Cycle and profitability of manufacturing firms of Bangladesh. Longer Cash Conversion Cycle indicates higher investment in inventories and debtors which in turn increases carrying cost and reduces profitability of the firm. Hence, firms having longer Cash Conversion Cycle should focus on shortening it by holding fewer days of production needs in raw material inventory, speeding up the production process, holding fewer goods in finished goods inventory, reducing the credit term offered to customers and taking longer time to pay trade creditors. However, the length of the Cash Conversion Cycle should be maintained at such a level that it does not lead to increase in shortage cost, which can also hampers firm's profitability.

## 6. Conclusion

This study presents the comparative average values of Cash Conversion Cycle for the six different manufacturing industries as listed in Dhaka Stock Exchange. Descriptive statistics reports that Cash Conversion Cycle varies from industry to industry. But the variation is not large enough to establish statistically significant differences among Cash Conversion Cycle of six manufacturing industries. However, several earlier studies have investigated the differences in working capital management across industries and also reported industry effect on firm's working capital policies.

Former studies of same kind showed that, firm size significantly affects cash conversion cycle. Jose et al (1996) found that larger firms tend to be more profitable and tend to have shorter Cash Conversion Cycle (as cited in Pedro et al., 2009). The findings of the present study also indicate that there exists a significant negative linear relationship between the Cash Conversion Cycle and the firm size, especially in terms of net sales. Therefore, small-scale manufacturing firms have longer Cash Conversion Cycle and vice-versa. The finding of the present study is also in line with the findings of Moss and Stine (1993). According to them, since longer Cash Conversion Cycles are associated with smaller firms this offers a strong incentive for these firms to better manage their Cash Conversion Cycle (as cited in Uyar, 2009)

The tradition link between Cash Conversion Cycle and firm's profitability is that shorter Cash Conversion Cycle reduces the time period for which cash is tied up in working capital, thus improving profitability. A longer Cash Conversion Cycle indicates high level of investment in working capital resulting in higher carrying costs for the firm which hampers profitability. Former studies by Wu (2001) and Chiou et al (2006) showed that the working capital requirement and the performance of the firm have effects on each other. In addition, these studies found that Return on Assets have a negative relationship with measures of working capital management (as cited in Pedro et al., 2009). The present study also shows significant negative linear relationship of Cash Conversion Cycle with Return on Equity in terms of manufacturing firms of Bangladesh. High opportunity cost or shortage cost is associated with shorter Cash Conversion Cycle whereas high carrying cost is associated with longer Cash Conversion Cycle. Therefore, the length of Cash Conversion Cycle can create both cost and benefits for the firm. Hence, understanding and weighting of these costs and benefits and efficient management of Cash Conversion Cycle is crucial for enhancing profitability of a firm.

## 7. Limitations and Scope for Future Research

This study comprises of only thirty manufacturing firms from six different manufacturing industries of Bangladesh. Since the sample size is small, this study could not establish the existence of significant differences in Cash Conversion Cycle among varying industries, which has been found in parallel studies.

Furthermore, being a study on a limited-scale, significant correlation between Cash Conversion Cycle and total assets and ROA, also could not be established. Therefore, future researches should include all the listed companies under all categories of manufacturing industries. Besides, the data obtained from the sample were only for the year 2008, which is one of the drawbacks of this study. However, future researches should include a large sample and cover a longer time period to investigate the differences in Cash Conversion Cycle among various manufacturing industries and also to determine its association with firm size and profitability.